\documentclass[%
 aip,
 sd,%
 amsmath,amssymb,
 reprint,%
]{revtex4-1}

\usepackage{graphicx}
\usepackage{dcolumn}
\usepackage{bm}

\begin{document}

\preprint{AIP/123-QED}

\title[Mechanical bound state in the continuum for optomechanical microresonators]{Mechanical bound state in the continuum for optomechanical microresonators}

\author{Yuan Chen}
\author{Zhen Shen}
\author{Xiao Xiong}
\author{Chun-Hua Dong}
\email{chunhua@ustc.edu.cn}
\author{Chang-Ling Zou}
\email{clzou321@ustc.edu.cn}
\author{Guang-Can Guo}
\affiliation{Key Laboratory of Quantum Information, Chinese Academy of
Sciences, University of Science and Technology of China, Hefei 230026, P. R. China;}
\affiliation{Synergetic Innovation Center of Quantum Information and Quantum Physics, University of Science and Technology of China, Hefei 230026, P. R. China.}

\date{\today}

\begin{abstract}
Clamping loss limits the quality factor of mechanical mode in the
optomechanical resonators supported with the supporting stem. Using
the mechanical bound state in the continuum, we have found that the
mechanical clamping loss can be avoided. The mechanical quality factor
of microsphere could be achieved up to $10^{8}$ for a specific radius
of the stem, where the different coupling channels between the resonator
and supporting stem are orthogonal to each other. Such mechanism is
proved to be universal for different geometries and materials, thus
can also be generalized to design the high quality mechanical resonators.
\end{abstract}

\maketitle

\section{Introduction}

The optomechanical interactions, which have been motivated more than
thirty years \cite{Ref_1,Ref_2,Opto_1,Opto_2}, take place via either
the radiation pressure and gradient forces induced by the optical
fields \cite{Radiation_1,Radiation_3,Li2008} or nonlinear processes
such as Brillouin scattering \cite{Brillouin_1,Brillouin_2}. These
interactions play a crucial role in the quantum state transfer
between light and mechanical motion \cite{Hybrid,Transducer2,Transducer1}.
The ubiquitous nature of mechanical freedom degrees can enable a macroscopic
mechanical oscillator to couple to nearly all types of quantum systems,
including charge, spin, atomic, and superconducting qubits, as well
as photons at nearly any wavelength \cite{Treutlein2014,Kurizki2015}.
The radiation-pressure-induced coherent mechanical oscillations in
whispering gallery mode (WGM) microresonator have been observed in
experiment about a decade ago \cite{Radiation_3}. Such optomechanical
resonators have been used to obtain a variety of coherent optical
processes, such as optomechanical induced transparency (OMIT) \cite{OMIT_1,OMIT_2,OMIT_3},
optomechanical light storage \cite{Storage}, and coherent optical
wavelength conversion \cite{OMIT_3,Conversion_1}. Radiation pressure
cooling of mechanical oscillators to their ground states has also
been realized experimentally in these optical resonators \cite{Cooling_1,Cooling_2}.
In all these applications, the intrinsic mechanical quality ($Q$)
factor is important, which limits the phonon lifetime and the fidelity
of interconversion between phonon and other excitations.
\begin{figure}[tp]
\includegraphics[ clip,width=8cm]{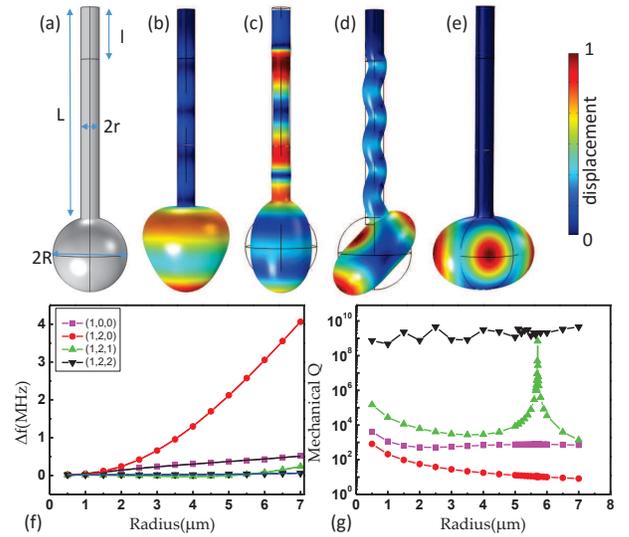}\caption{Mechanical modes in the microsphere. (a) The microsphere with the
supporting stem. (b) The radial breathing mode $(1,\:0,\:0)$ with
frequency of $127.10$MHz and mechanical $Q=721$. (c) $(1,\:2,\:0)$
mode with frequency of $89.20$MHz and $Q=15$. (d) $(1,\:2,\:1)$
mode with frequency of $87.48$MHz and $Q=3874$. (e) $(1,\:2,\:2)$
mode with frequency of $87.53$MHz and $Q=1.2\times10^{8}$. The parameters
$(R,\:r,\:L,\:l)$ of the silica microsphere are $(18,\:4.5,\:105,\:25)\mathrm{\mu m}$.
The color represents the displacement of mechanical vibration. (f)
The frequency drift versus the radius of the stem, for fixed sphere
size. (g) The mechanical $Q$ versus the radius of the stem, for fixed
sphere size. }
\end{figure}

The dominate origins of mechanical energy dissipation can be divided
into three categories: fluid-structure interactions, material damping
and clamping damping \cite{Cleland,limited_1}. There exists different
strategies to reduce the relevant dampings. To eliminate the
fluid-structure interactions or air damping, the most effective strategy
is to place the device in the low pressure vacuum chamber \cite{Limited_2}.\textbf{
}For the material damping, controlling the type, distribution, density
of defects, the operating temperature, and designing material are
usually considered \cite{limited_1}. Although the material damping
face the fundamental technique challenges, the clamping damping can
be eliminated by clever designs, such as supporting the resonator
at its nodal points \cite{limited_6,limited_9} or surrounding the
resonator by phononic band-gap structures \cite{limited_13}.

Here, we propose the mechanical bound state in the continuum (BIC)
to reduce the clamping damping to show the improved mechanical $Q$
in the silica microsphere. The BIC, which was first proved by Von
Neuman and Wigner in 1929 \cite{BIC_1}, has been demonstrated in electronic
and photonic structures \cite{BIC_3,BIC_4,BIC_5,BIC_6,Hus,BIC_2}.
Similar to the optical BIC that creates two dissipation channels and
controls their phases to cancel each other \cite{BIC_2}, the elastic
wave dissipation through different polarizations can be eliminated
in appropriate structure. By extending the BIC phenomena to the mechanical
system, we demonstrate the leakage of spherical vibration to the supporting
stem can be canceled. The mechanical Q factor limited by the clamping
loss can be as high as $10^{8}$ (limited by the precision of simulation)
for BIC with the appropriate supporting stem. We believe that the
BIC is a novel and efficient method for inhibiting the mechanical
clamping loss, the mechanism is universal and can be used for the designing of other phononic
structures to overcome the phonon dissipation.

\section{Model}

The basic principle of the mechanical
BIC is illustrated by the simple microsphere on stem structure (Fig.$\,$1(a)). In
the experiment, the silica microsphere is normally fabricated by a
$\mathrm{CO_{2}}$ laser beam and supported at the end of the fiber
tip \cite{SphereReview}. In past decades, the silica microsphere
on fiber stem has been extensively studied in experiment due to its
ultrahigh optical $Q$ factor (up to $10^{10}$ \cite{high_q,high_Q})
and easy fabrication. Thus, we build the model in the COMSOL multiphysics,
as shown in Fig.$\,$1(a). The microsphere with radius of $R$ is
supported by a stem with radius of $r$. The stem length is $L$ with
a perfectly matched layer of $I$. The parameters of the model are
$(R,\:r,\:L,\:I)$=$(18,\:4.5,\:105,\:25)\mathrm{\mu m}$. Similarly
with the optical modes \cite{SphereReview}, the acoustic waves in
silica microsphere can be identified by radial ($q$), orbital (\emph{$l$}),
and azimuth ($m$) numbers. In bulk materials, there are three types of
elastic wave for different polarizations, including pressure wave
(displacement along propagation direction), flexural wave (out-of-plane
transverse displacement) and shear wave (in-plane transverse displacement)
\cite{Cleland}. In spheres, the modes are divided into two categories\cite{limited_0}:
spheroidal and torsional modes, which are hybridized flexural and
pressure waves and pure shear waves, respectively. For optomechanical
applications, we are mostly interested in those spheroidal modes since
the volume change of torsional mode is zero. In addition, it's also
recently demonstrated that the spheroidal mode can couple with magnons
efficiently \cite{magnon}.

Shown in Figs.$\,$1(b)-(e) are the typical spheroidal modes ($1,0,0$),
($1,2,0$), ($1,2,1$) and ($1,2,2$), respectively. From the simulated
distributions of deformation, we find that there are three different
clamping conditions: the spherical vibration coupling with the pressure
wave in the supporting stem for ($1,0,0$) and ($1,2,0$) modes {[}Figs.
1(b) and (c){]}, the excited flexural wave in the stem for ($1,2,1$)
{[}Fig. 1(d){]}, and almost no elastic wave dissipation in the stem
for ($1,2,2$) mode. The intuitive explanation of the excited pressure
wave for the $\left(q,l,0\right)$ is that the motion of the sphere
is at the polar point (antinode), or pure radial motion at the clamping
point, which can only compress and stretch the stem in $z$-direction
(cylindrical coordinate for stem). In contrast, for ($q,l,m$) mode
with $m=l$, there is only vibration along the equator of the
sphere (we may call this type of mode as mechanical WGMs), so there
is no displacement at the polar point, then the loss is zero. When
varying the stem size, we can see that the frequencies of ($1,0,0$)
and ($1,2,0$) modes change a lot while the mechanical $Q$ factors
decrease monotonously. It indicates the coupling between stem and
these modes proportional to the area of the cross-section of the stem ($\pi r^{2}$).
For the mechanical WGMs, both the frequency and mechanical $Q$ remain
constant for different $r$, respectively, and the mechanical $Q$
is larger than $10^{8}$ with small fluctuations due to the numerical
errors.

The ($1,2,1$) mode shows very different and interesting behavior
from the other two type of modes. From the field distributions, we
can see that at the contact point of the sphere and the stem, there
is zero mean displacement in $z$-direction but non-zero displacement
in transverse direction. The flexural motion (out-of-plane transverse
displacement) of stem shows similar displacement in $z$-direction
and transverse direction. Shown in Fig. 1(g), there is a singular
peak for ($1,2,1$) mode at $r\approx5.6\,\mathrm{\mu m}$. At the
peak, the mechanical $Q>10^{8}$ with the number limited by the precision
of numerical simulation.

\begin{figure}[tp]
\includegraphics[ clip,width=7cm]{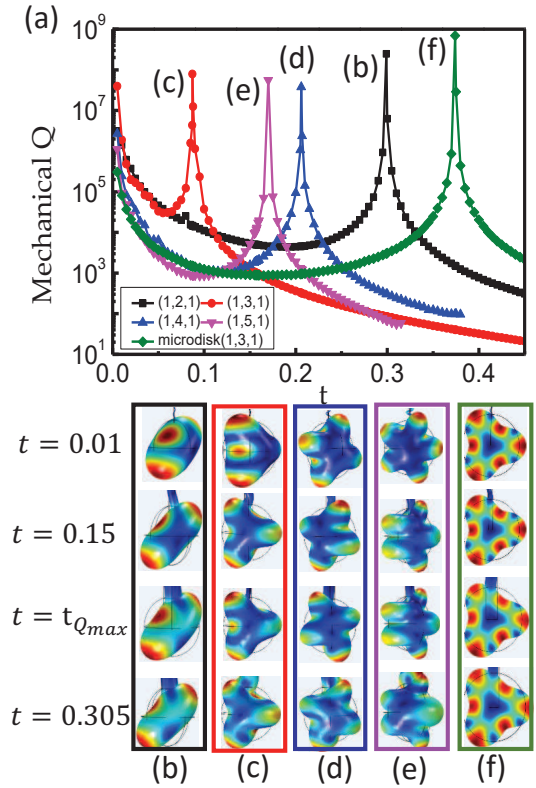}\caption{Universality of the mechanical Q extremum of five mechanical systems
versus the ratio $t$ of $r/R$. (a) The black, red, blue, magenta lines show the ($1,2,1$),
($1,3,1$), ($1,4,1$), ($1,5,1$) modes with fixed sphere radius
$R=12\,\mathrm{\mu m}$. The green line shows the ($1,3,1$) mode
with the quasi-2D microdisk ($R=20\,\mathrm{\mu m}$) supported by
the beam. (b)-(f) The legends in black, red,
blue, magenta, green-colored boxes are the displacements of the mechanical
systems with $t=0.01,\:0.15,\:t_{Q_{max}},\:0.305$. The corresponding
($t_{Q_{max}}$, $Q_{max}$) are ($0.299,\:2.5\times10^{8}$), ($0.087,\:7.9\times10^{7}$),
($0.206,\:3.8\times10^{7}$), ($0.17,\:5.6\times10^{7}$) , ($0.374,\:6.9\times10^{8}$),
respectively. }
\end{figure}

\section{Mechanical BIC}

For the purpose of optomechanical interaction in the sphere, we would
choose the two types of modes, ($1,2,1$) in Fig.$\,$1(d) and ($1,2,2$)
in Fig.$\,$1(e), in our experiments. The ($1,2,2$) mode shows zero
net displacement in the equator and non-zero orbit angular momentum,
thus can only couple with light through the Brillouin scattering process
\cite{Brillouin_1,Brillouin_2}, which requires two optical modes
to satisfy the phase matching and energy conservation relations. The
mechanical loss of the breath type of modes, ($1,0,0$) in Fig.$\,$1(b) and ($1,2,0$)
in Fig.$\,$1(c), are very large due to the
strongest radiation pressure coupling and the pressure wave leakage
to the stem. To achieve high mechanical Q factor for those modes, the supporting stem should be thinner than $1\,\mathrm{\mu m}$. However, the system would be very fragile and impractical for thin stems, the ($1,2,1$) mode is a better choice with much high mechanical Q factor and thick stem size. The singular
high-Q peak in the ($1,2,1$) mode is useful and the details about
the mechanism of the phenomena are explained in the following.

In Fig.$\,$2(a), the mechanical Q of ($1,2,1$) mode profiles with
different stem to sphere radius ratio $t=r/R$. At the singular point,
the mechanical Q exponentially decays with the deflection of the optimal
ratio. This extremum of the mechanical Q means that the propagating
flexural wave in the stem can not be excited by the sphere vibration.
We notice that there are two polarizations of displacement involved
into the coupling between the spheroidal mode of sphere and flexural
mode of stem: the radial displacement and tangent displacement (in
respect to the sphere). The polarization of the sphere combines the
two displacements and it is the same with the stem. Therefore, it
is possible that the polarization of the sphere from one channel is
orthogonal to the polarization of the stem from the other channel
in the joint, which realizes the mechanical BIC and forbids the elastic
wave leakage to the continuum waves of stem.

Similar principle should also work for other mechanical modes, such
as all the higher order $(1,l,1)$ modes. To verify such mechanical
BIC mechanism in the spherical mechanical microresonator, the higher
order $(1,l,1)$ modes with $l=3,4,5$ are studied and the results
are plotted in Figs.$\,$2(a) and (c)-(e). All these modes show similar
singular high mechanical Q factors with specific $t$. At those optimal
$t$, the displacement of stem exponentially decreases with the distance
to the sphere, which indicates that the propagating elastic wave in
the stem is not excited. We should note that the BIC for ($1,3,1$)
is not supported in the silica sphere, since the spheroidal mode couples
to the high loss torsional mode by the perturbation of stem. Therefore,
we modified the Young's modulus to 50$\,$GPa and the Poisson ratio
to 0.12, and plotted the results in Figs. 2(a).

Then, we testify the existence of the BIC for different material properties.
As shown in Fig.$\,$3, the mechanical $Q_{max}$ and the eigenfrequency
for $(1,2,1)$ mode are changed according to Young's modulus and Poisson
ratio of the material. Firstly, for R=12$\,\mathrm{\mu m}$, the eigenfrequency
increases with the Young's modulus rising from $50$ to $100$ GPa.
The mechanical $Q_{max}$ for BIC retains up to $10^{8}$. Due to
the precision limited by numerical simulation, the fluctuation of
the mechanical $Q_{max}$ above $10^{8}$ is unconspicuous. We find
that the optimal ratio $t$ is stable as the Young modulus of the
structure adds up. As shown in Figs. 3(c) and (d). When varying the
Poisson ratio from $0.11$ to $0.25$, the eigenfrequency decreases,
and $t$ rises up while the mechanical $Q_{max}$ remains high.

\begin{figure}[tp]
\includegraphics[clip,width=8cm]{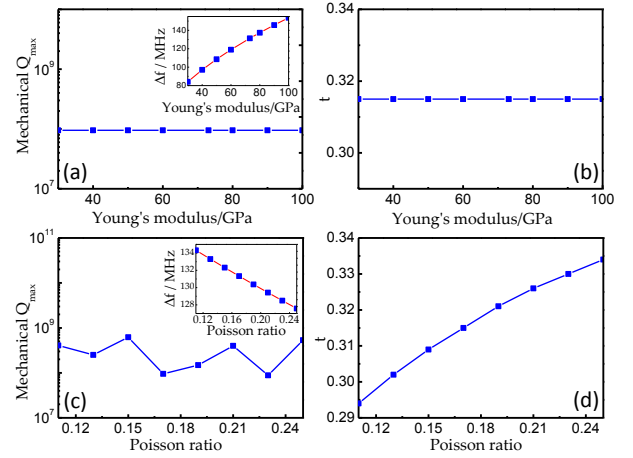}\caption{The influence of the eigen frequency, the optimal t and the mechanical
$Q_{max}$ by material. (a-b) The mechanical $Q_{max}$, and the optimal
rate $t$ versus the Young's modulus of the microsphere with the stem,
rising from $50$GPa to $100$GPa. The inset in (a) shows the eigen
frequency versus the Young's modulus from 50GPa to 100GPa. (c-d) The
mechanical $Q_{max}$ , and the optimal $t$ versus the poisson ratio
of the structure varying from $0.11$ to $0.25$. The inset in (c)
shows the eigen frequency versus the poisson ratio from $0.11$ to
$0.25$.}
\end{figure}

Further generalization of such principle can apply to the quasi-2D
structure, such as the microdisk suspended by attaching to an auxiliary
beam\cite{ring resonator}. Similar to the microsphere, the 3rd order
mode in Figs.$\,$2(f) (similar to spherical coordinator, this mode
in disk can be identified as ($1,3,1$) in respect to the mode order
in the radial, azimuthal and thickness direction) shows in-plane standing
wave profile along the boundary of the disk. The mode is combined
radial and tangential displacements, and the corresponding leaking
wave in beam is the in-plane flexural wave. Very similar to the microsphere
case, we observe the leakage forbidden by changing the beam width, which
leads to the BIC phenomenon in structures on photonic chips.

\section{Discussions}

All of these results confirm the universality of the mechanism of
mechanical BIC, which is applicable to other material and structures.
For practical applications of the mechanical BIC, we should also consider
the imperfections of the fabricated microstructures. For example,
during the fabrication of the silica microsphere, due to the surrounding
environment perturbations (such as wind and vibrations) and gravitation
force, the sphere may not be perfectly spherical and the stem may
not point to the center of sphere.

\begin{figure}[tp]
\includegraphics[clip,width=8cm]{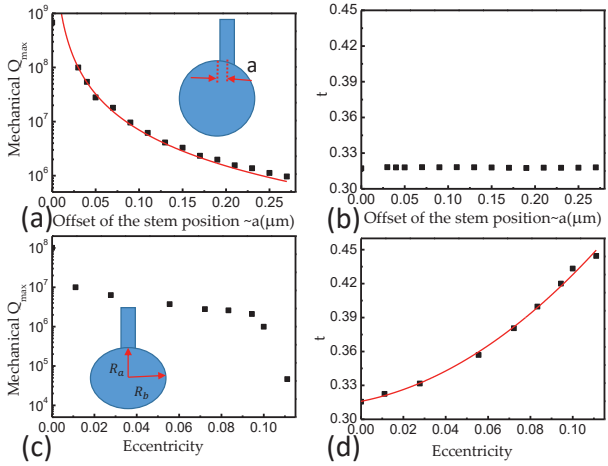}\caption{(a-b) The mechanical $Q_{max}$ and the optimal rate $t$ versus the
offset of the supporting stem, respectively. (c-d) The mechanical
$Q_{max}$ and optimal rate $t$ versus the eccentricity of the microresonator,
respectively. The fitted curves of (a) and (d) are polynomial fititngs. }
\end{figure}

Figs.$\,$4(a) shows that the achievable mechanical $Q_{max}$ for
BIC reduces with the increased offset of the stem. When the offset
is $0.25\,\mathrm{\mu m}$ for $R=18\,\mathrm{\mu m}$, the mechanical
$Q_{max}$ value limited by the clamping loss gets down to $10{}^{6}$.
Therefore, the symmetry is important for the high mechanical $Q_{max}$.
In Figs.$\,$4(b), the optimal $t$ is almost constant for different
offsets. For the gravity induced small eccentricity of the microsphere,
the achievable mechanical $Q_{max}$ also reduces, as shown in Figs.$\,$4(c).
Define the eccentricity as $\frac{R_{b}-R_{a}}{R_{b}}$, with $R_{b}$
and $R_{a}$ are long and short axes. There is a critical eccentricity
around $0.1$, beyond that the mechanical $Q_{max}$ reduces dramatically
to be smaller than $10^{5}$. The results in Figs.$\,$4(d) indicate
that the optimal rate $t$ rises quickly as the eccentricity grows
up. In practical experiment, the gravity induced eccentricity is usually
around $0.02\sim0.03$, which still permits $Q_{max}$ be larger than
one million.

In previous experiments, the measured mechanical Q of silica microsphere
is limited to be $2\times10^{4}$ in vacuum and cryogenic temperatures
\cite{Cooling_1,Cooling_2}. It is believed that it is the material
limitation, due to the absorption of two-level system (TLS) in non-crystal
silica. However, such limitation can be resolved by pump the system
to saturate the TLS at low temperature \cite{TLS}. By this technique,
the million level $Q$ of mechanical BIC is possible for silica whispering
gallery microresonators and will promote the single photon level optomechanics
experiments. For example, the lifetime for mechanical mode with $Q=10^{6}$
and $f=10^{8}$Hz is $\tau\approx1.6\,\mathrm{ms}$, which is one
excellent candidate for quantum memory.

\section{Conclusion}

The mechanical bound states in the continuum are studied, whose elastic
energy loss through clamping structures is forbidden. The intrinsic
mechanism is that the polarization of the vibration on the mechanical
resonator is orthogonal to that of the supporting stem, thus cancels
the elastic wave radiation through changing the radius of the stem.
The detailed studies on different mechanical modes, different mechanical
resonator structures and materials confirm the universality of the
mechanical bound states in the continuum. Therefore, the mechanism
studied in this paper can be used for the design of other mechanical
resonators, in the field of optomechanics and electromechanics.

\section*{Acknowledgment}

CLZ thanks H. Wang for stimulating discussions. The work was supported
by the Strategic Priority Research Program (B) of the Chinese Academy
of Sciences (Grant No. XDB01030200), National Basic Research Program
of China (Grant Nos. 2011CB921200 and 2011CBA00200) and the National
Natural Science Foundation of China (Grant Nos. 61308079 and 61505195),
Anhui Provincial Natural Science Foundation (Grant No. 1508085QA08),
the Fundamental Research Funds for the Central Universities.

\end{document}